\newif\ifpreprint
  \preprintfalse        
  \preprinttrue         

\ifpreprint
	\documentstyle[aas2pp4,epsf,twoside,fleqn]{article}
	\newcommand{\fL}[1]{{\footnotesize\bf\uppercase{#1}}}
	\newcommand{\nc}[1]	{{\,{#1}\,}}
\else
 	\documentstyle[12pt,aasms4,epsf]{article}
	\newcommand{\nc}[1]	{#1}
\fi

\newcommand{\bea}	{\begin{array}}
\newcommand{\eea}	{\end{array}}
\newcommand{\beq}	{\begin{equation}}
\newcommand{\eeq}	{\end{equation}}
\newcommand{\ben}	{\begin{eqnarray}}
\newcommand{\een}	{\end{eqnarray}}
\newcommand{\bsq}	{\begin{mathletters}}
\newcommand{\esq}	{\end{mathletters}}

\newcommand{\eq}	{\nc{=}}	
\newcommand{\mi}	{\nc{-}}	
\newcommand{\pl}	{\nc{+}}	
\newcommand{\id}	{\nc{\equiv}}	
\newcommand{\ap}	{\nc{\approx}}	

\newcommand{\B}[1]	{\mbox{\boldmath$#1$}}
\newcommand{\p}		{\partial}
\newcommand{\D}		{{\rm d}}
\newcommand{\pr}	{\prime}
\newcommand{\JR}	{J_R}
\newcommand{\oR}	{\omega_R}
\newcommand{\half}	{\case{1}{2}}
\newcommand{\ov}[1]	{\overline{#1}}
\newcommand{\fsub}[1]	{{f_{\rm #1}}}
\newcommand{\fshu}	{\fsub{Shu}}
\newcommand{\fnew}	{\fsub{new}}
\newcommand{\fa}	{\fsub{a}}
\newcommand{\fb}	{\fsub{b}}

\newcommand{\eqn}[1]	{equation\ (\ref{#1})}
\newcommand{\Sec}[1]	{Section\ \ref{sec:#1}}
\newcommand{\Par}[1]	{\S\ref{sec:#1}}
\newcommand{\Fig}[1]	{Figure\ \ref{fig:#1}}
\newcommand{\fig}[1]	{Fig.\ \ref{fig:#1}}

\ifpreprint
  \slugcomment{scheduled for publication in the September 1999 issue of
		The Astronomical Journal}
\else
  \lefthead{Walter Dehnen}
  \righthead{Distribution functions for disks}
\fi

\begin{document}

\ifpreprint \thispagestyle{empty} \fi
\title{Simple Distribution Functions for Stellar Disks}
\author{Walter Dehnen\altaffilmark{1}}
\affil{Theoretical Physics, 1 Keble Road, Oxford OX1 3NP, United Kingdom}
\altaffiltext{1}{e-mail:	dehnen@physics.ox.ac.uk}

\begin{abstract} \ifpreprint\noindent\fi
Distribution functions (DFs) for dynamically warm thin stellar disks residing 
in arbitrary axisymmetric potentials are presented which approximately reproduce
pre-described surface-density and velocity-dispersion profiles. The functional 
form of the DFs is obtained by `warming-up' a model made entirely of circular 
orbits. This can be done in various ways giving different functional forms for
the DF. In the best case, the DF reproduces the pre-described profiles to 
within a few per cent for a typical case reminiscent to the old stellar disk in 
the Milky Way. This match may be improved to about one per cent or better by a 
simple iterative method. An algorithm is given to draw phase-space points 
randomly from the DFs for the purpose of, e.g., $N$-body simulations. All the 
relevant computer programs are available from the author.
\end{abstract}


\ifpreprint	\section{I\fL{ntroduction}}\label{sec:intro}\noindent
\else		\section{Introduction}\label{sec:intro}
\fi
All the dynamical properties of a collisionless stellar system are prescribed
by its phase-space distribution function (DF) in conjunction with the underlying
gravitational potential $\Phi$. The knowledge of a {\em simple\/} DF is often 
vital for studies of the dynamical state of stellar systems, their equilibrium 
properties and time-evolution. In this paper, simple DFs for thin axisymmetric 
disks are given. These DFs may be used in studies of, e.g., spiral structure, 
the effects of non-axisymmetric perturbation on the velocity distribution in the
solar neighborhood, and to seed initial conditions for $N$-body simulations.

For a thin axisymmetric disk, $\Phi\eq\Phi(R)$ and the collisionless Boltzmann 
equation tells us that in dynamical equilibrium the stellar DF must be a 
function of the stellar energy and angular momentum alone, $f\eq f(E,L)$. 
Thus, compared to the more general case of three-dimensional axisymmetric or 
non-axisymmetric systems, where the DF may also depend on non-classical 
integrals of motion, the DFs of thin axisymmetric disks are simple and allow 
for mathematical manipulations. It is possible to give a DF $f(E,L)$ which 
implements some pre-defined properties, for example, a certain surface density 
profile $\Sigma(R)$, in a given underlying gravitational potential $\Phi(R)$%
\footnote{
        In this paper, we will not restrict ourselves to {\rm self-consistent\/}
        models in which the gravitational potential is itself generated by the 
        surface density given through the DF, but allow for a general $\Phi$, 
	which may partly be due to the stellar disk.}.
Unfortunately however, this DF is highly non-unique: every one of the infinitely
many ways to expand $\Sigma(R)$ as a function of both $R$ and $\Phi(R)$, and 
thus to obtain a so-called surface-density partition $\Sigma(R,\Phi)$, 
determines uniquely the part of the DF even in $L$ (\cite{kal76}). This even
part of the DF, $f_+$, can be computed (usually numerically) by a path-integral 
(\cite{hq93}). Additional constraints, for example, a certain 
radial-velocity-dispersion profile $\sigma_R(R)$, restrict the possible 
surface-density partitions and thus DFs, but do not remove the general 
degeneracy (obviously, the DF being a function of two variables cannot be 
uniquely determined by a finite set of functions of one variable). 

Given this degeneracy among possible disk DFs, there are two different 
ways to construct specific dynamical models. The first approach follows
the above lines in specifying a particular surface-density partition and 
then solving for the implied $f_+(E,L)$. This technique has the advantage that 
the pre-defined properties are {\em exactly\/} satisfied by the resulting DF, 
but the disadvantage that the DF can rarely be given in closed form, which
tends to be complicated.

The second way to a disk DF is to specify a particular functional form,
that results only in {\em approximate\/} agreement with the pre-defined 
properties. This approach has the advantage that the DF is always of a simple 
functional form, which often is of great importance in dynamical modeling. 
Another bonus is that the degeneracy among DFs with given surface density 
and velocity dispersion is removed by choosing a particular functional form for
the DF itself rather than the partition function. This enables one to choose
the functional form for $f(E,L)$ on the basis of astro-physical arguments.
On the other hand, in the first approach the degeneracy is often exploited by
choosing the surface-density partition such that the resulting DF is analytic.

The aim of this paper is to improve and extend on what has been done before
on that second approach to warm-disk DFs. In \Sec{warm-dfs}, warm-disk DFs are
constructed by warming-up the DF for a completely cold disk in which all stars 
on circular orbits. In \Sec{assess}, the ability of four types of warm-disk DFs 
to reproduce the predefined surface-density and velocity-dispersion profile is 
assessed and a simple algorithm is given to improve on this ability. \Sec{kinem}
compares the velocity moments up to fourth-order for the four DFs introduced in 
\Sec{warm-dfs}, while \Sec{sample} gives an algorithm for drawing from the DFs 
samples of phase-space points such as might be used for numerical simulations.

\ifpreprint	\section{R\fL{ecipes for} W\fL{arm}-D\fL{isk} D\fL{istribution}
			 F\fL{uNctions}} \label{sec:warm-dfs}\noindent
\else		\section{Recipes for Warm-Disk Distribution Functions}
		\label{sec:warm-dfs}
\fi
We seek DFs $f(E,L)$ which approximately create given {\em targets\/} for the
surface density $\Sigma(R)$ and radial velocity dispersion $\sigma_R(R)$ in a 
given potential $\Phi(R)$. The basic technique is to modify the DF for a 
completely cold disk.

\subsection{The Distribution function for the cold disk} \label{sec:cold-df}
\ifpreprint\noindent\fi
Let $E_c(L)$ be the energy of the circular orbit with angular momentum $L$
(see, e.g., Appendix A of Dehnen 1999, hereafter paper~I, for details of the 
properties of circular orbits), then
\beq\label{Shu-ansatz} 
	\fsub{cold}(E,L) = \fsub{circ}(L) \;\delta\big(E-E_c(L)\big),
\eeq
describes a DF which allows for circular orbits only. From 
\beq\label{M}
	M = (2\pi)^2 \int\D^2\!\B{J}\; f,
\eeq
which may be derived using $\D^2\!\B{x}\,\D^2\!\B{v}\equiv\D^2\!\B{J}\,\D^2\!
\B{\theta}$ with $\B{J}\equiv(J_R,L)$ denoting the action variables, and
$(\p E/\p J_R)_L\equiv\oR$, we have
\beq \label{Mel}
	M = (2\pi)^2 \int \D E \int \D L\;f(E,L)\;\omega_R^{-1}(E,L),
\eeq
and thus
\beq \label{dMel}
	\D M = (2\pi)^2 \fsub{circ}(L)/\kappa(L)\; \D L,
\eeq
where I have used the fact that on circular orbits the radial frequency $\oR$ 
equals the epicycle frequency $\kappa$. On the other hand we have
\beq \label{dMR}	
	\D M = 2\pi\, R\; \Sigma(R) \,\D R,
\eeq
and equating (\ref{dMel}) with (\ref{dMR}) gives, with $\D L_c/\D R\eq R\kappa/
\gamma$ (cf.\ equations A2 to A6 of paper~I), $\fsub{circ}(L)\eq f_1\big(R_c(L)
\big)$ with
\beq \label{fone}
	f_1(R) \equiv {\gamma(R)\Sigma(R)\over2\pi}.
\eeq
Here, $R_c(L)$ is the radius of the circular orbit with angular momentum 
$L$, while $\gamma\id2\Omega/\kappa$ with the circular frequency $\Omega(R)$.

\subsection{The Shu distribution function} \label{sec:shu-df}
\ifpreprint\noindent\fi
In order to allow for non-circular motions, the $\delta$-function in the
cold-disk DF (\ref{Shu-ansatz}) must be replaced by some function of finite 
width. Since the epicycle energy $E-E_c(L)$ is non-negative, a simple choice 
is an exponential corresponding to a Maxwellian velocity distribution
\beq \label{exp}
	\delta\big(E-E_c(L)\big)\;\longrightarrow\; {1\over\sigma_R^2}
	\exp\!\left[{E_c(L)-E\over\sigma_R^2}\right].
\eeq
Using $R_L\id R_c(L)$ also as argument of $\sigma_R^2$, this leads to the
well-known DF (Shu 1969)
\beq \label{fshu}  
	\fshu(E,L) = {\gamma(R_L)\,\Sigma(R_L)\over2\pi\,\sigma_R^2(R_L)}
			\;\exp\!\left[{E_c(L){-}E\over\sigma_R^2(R_L)}\right].
\eeq

\subsection{New warm-disk distribution functions} \label{sec:new-df}
\ifpreprint\noindent\fi
Instead of (\ref{Shu-ansatz}), we could have started with the alternative ansatz
for a cold-disk DF:
\beq\label{New-ansatz} 
	\fsub{cold}(E,L) = \fsub{circ}(E) \;\delta\Big(\Omega(E)
					\big[L_c(E)-L\big]\Big).
\eeq
With $\D E_c/\D R\eq R\kappa^2/2$, one finds $\fsub{circ}(E)=f_1\big(R_c(E)\big
)$, where $R_c(E)$ is the radius of the circular orbit at energy $E$. As 
$\fshu$ was derived from (\ref{Shu-ansatz}), we may derive a new warm-disk DF 
from (\ref{New-ansatz}):
\beq\label{fnew}
	\fnew(E,L) = {\gamma(R_E)\,\Sigma(R_E)\over2\pi\,\sigma_R^2(R_E)}\,
		\exp\!\left[{\Omega(R_E)[L{-}L_c(E)]\over\sigma_R^2(R_E)}\right]
\eeq
with $R_E\id R_c(E)$.
There is actually large freedom in designing more warm-disk DFs by combining
the various possibilities for the arguments of $f_1$, $\sigma_R^2$, and the 
exponential. Two other possible DFs are
\ben
\label{fa} \fa(E,L) &=& {\gamma(R_E)\,\Sigma(R_E)\over2\pi\,\sigma_R^2(R_E)}\;
	\exp\!\left[{E_c(L){-}E\over\sigma_R^2(R_E)}\right],		\\[1ex]
\label{fb} \fb(E,L) &=& {\gamma(R_L)\,\Sigma(R_L)\over2\pi\,\sigma_R^2(R_L)}\;
	\exp\!\left[{\Omega(R_L)[L{-}L_c(E)]\over\sigma_R^2(R_L)}\right].
\een

\subsection{Physical Motivation} \label{sec:motiv}
\ifpreprint\noindent\fi
Shu (1969) motivated his DF by the idea that an initial process of violent 
relaxation formed the distribution function of stellar disks. He also assumed 
that the initial distribution of angular momenta is not changed by this process.
However, we now know, that these ideas are incorrect. Disk stars gain their 
random motions by scattering off molecular clouds, spiral waves, satellite 
galaxies and other agents. Therefore, the velocity dispersion grows 
continuously, which is also reflected in the observed increase of random motions
with stellar age in the solar neighborhood. Such scattering processes do not 
change the value of a star's Hamiltonian in the frame co-rotating with the 
scattering agent. This Hamiltonian is equal to the Jacobi energy $E_J\eq E\mi
\omega L$ with $\omega$ the angular frequency of the scattering agent. 
Consequently, both the stellar energies and angular momenta are affected by 
disk heating. Thus, we may describe the DF as a warmed-up version of a DF made 
from new-born stars. Stars are born already with some finite velocity 
dispersion, which is small compared to that gained subsequently, and we may 
safely assume the DF of new-born stars is dynamically completely cold. The 
warming-up may be described by an exponential in the radial action $J_R$ 
(\cite{bin87})
\beq \label{f-motiv}
	f 	\approx f_1(R_1)\; {1\over\sigma_R^2(R_1)}\,
		\exp\!\left[-{\oR\JR\over\sigma^2_R(R_1)}\right].
\eeq
Here, $R_1\eq R_1(E,L)$ is some measure of the mean orbital radius that ensures 
the prescribed surface density to be closely matched by the DF. The radial 
action may be estimated from the classical epicycle theory to be
\[
	\oR\JR \sim    E - E_c(L),
\]
but, as demonstrated in paper~I,
\[
	\oR\JR \approx \Omega(E) \big[L_c(E)-|L|\big]
\]
is much better an approximation for near-flat rotation curves. Using either of
these approximations for $\oR\JR$ and either $R_L$ or $R_E$ for $R_1$, 
any of the four DFs presented above can be derived from \eqn{f-motiv}.

\ifpreprint 
  \begin{figure*}[t]
	\centerline{ \epsfxsize=140mm\epsfbox[27 325 590 716]{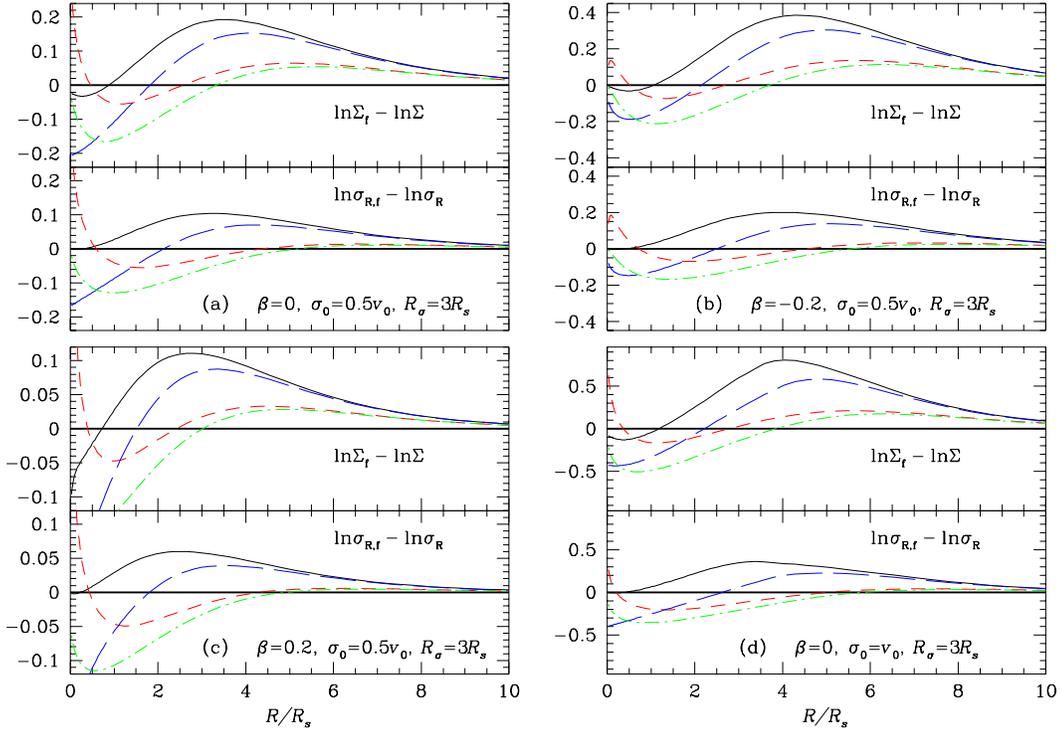}}
	\caption[]{\footnotesize
	Logarithmic deviations between the target surface density and velocity 
	dispersion ($\Sigma$, $\sigma_R$) and those actually created by the 
	warm-disk DF ($\Sigma_f$, $\sigma_{R,f}$) in power-law potentials
	with circular speed $v_c\nc\propto R^\beta$. Solid lines are for 
	$\fshu$, dashed for $\fnew$, dot-dash for $\fa$, and long-dashed for 
	$\fb$. In all four sub-figures, both $\Sigma$ and $\sigma_R$ decay
	exponentially with scale lengths $R_\sigma\eq3R_s$. In (a), (b), 
	and (c), the central velocity dispersion is $0.5v_0$, while it is twice 
	as large for (d). The rotation curve is flat in (a) and (d), slightly
	falling in (b) and slightly rising in (c). Notice the different scales
	of the $y$-axes. \label{fig:comp} }
  \end{figure*}
\fi

There are two arguments in favor of $\fnew$ being the most useful of these 
four DFs. First, the mean orbital radius $\ov{R}(E,L)$ is much better 
approximated by $R_E$ than by $R_L$ (in general: $R_L\nc\le R_E\nc\la\ov{R}$). 
Similarly, $\kappa(R_E)$ approximates the radial frequency $\oR$, which we 
replaced by $\kappa$ in the derivation of $f_1$, far better than $\kappa(R_L)$, 
see, e.g.\ Fig.~8 of paper~I. Therefore, one expects DFs with $R_1\eq R_E$ to 
create surface density and velocity dispersion in better agreement with the 
target functions than DFs with $R_1\eq R_L$. Second, $\Omega(L\mi L_c)$ as 
argument of the exponential naturally extends to negative $L$, while an 
exponential in $E\mi E_c$ does not allow for this possibility. This is 
significant insofar, as the tail of stars with small negative $L$ is known to 
be important for supporting disks against dynamical instabilities.

\ifpreprint 
  \begin{figure}[t]
	\centerline{ \epsfxsize=70mm\epsfbox[25 325 303 716]{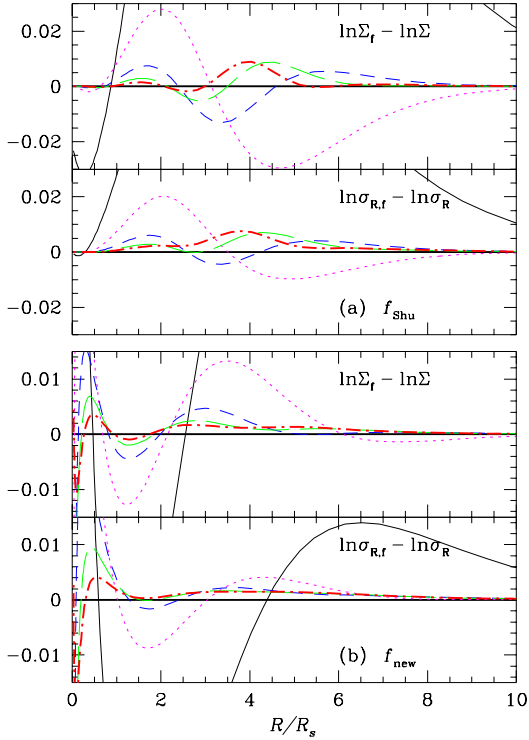} }
	\caption[]{\footnotesize
	The same as in \fig{comp}a, but for the DFs $\fshu$ (a) and $\fnew$ (b)
	after none (thin), one (dotted), two (dashed), three (long-dashed), and 
	four (bold dot-dash) iterations of the scheme of \Par{algol}. Notice 
	the different scales on the $y$-axes. \label{fig:iter} }
  \end{figure}

  \begin{figure}[t]
	\centerline{ \epsfxsize=70mm\epsfbox[25 325 303 716]{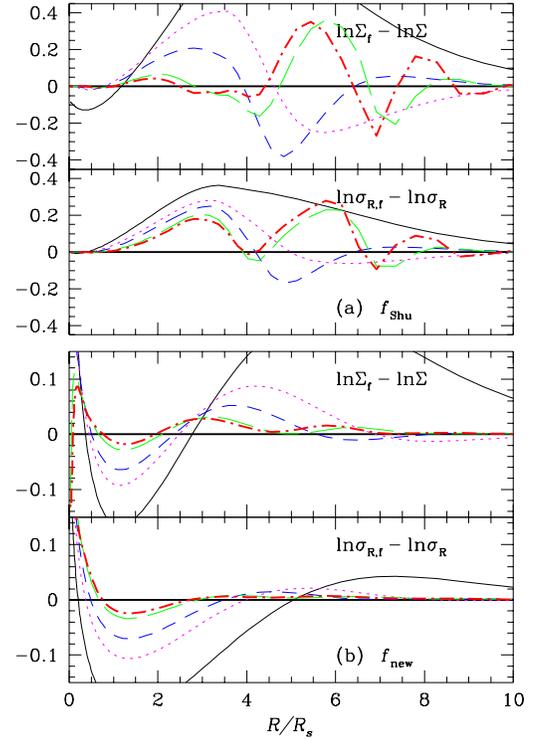} }
	\caption[]{\footnotesize
	As \fig{iter} but for the stellar disk and rotation curve as in 
	\fig{comp}d (which is hot inside $\sim1.3R_\Sigma$) and after none 
	(thin), one (dotted), two (dashed), five (long-dashed), and seven (bold 
	dot-dash) iterations of the scheme of \Par{algol}. Notice the different 
	scales on the $y$-axes. \label{fig:iter-two} }
  \end{figure}
\fi
\ifpreprint	\section{A\fL{ssessing the} D\fL{istribution} F\fL{unctions}}
		\label{sec:assess}\noindent
\else		\section{Assessing the Distribution Functions}
		\label{sec:assess}
\fi
Allowing for non-circular motions, i.e.\ replacing the $\delta$-function in 
(\ref{Shu-ansatz}) or (\ref{New-ansatz}) with some function of finite width, 
leads at any position $R$ to a mixture of stars originating from different radii
$R_1$. As a consequence, the surface density $\Sigma_f(R)$ that is actually 
created by the DF deviates from the target $\Sigma(R)$. Similarly, the radial 
velocity dispersion $\sigma_{R,f}(R)$ due to the DF is expected to differ from 
the target $\sigma_R(R)$. On the one hand a simple and generally applicable 
functional form for the DF is very desirable, on the other hand it is equally 
important for the DF to re-produce these targets as closely as possible. 
In \Par{comp}, we assess the ability of the DFs introduced in \Par{warm-dfs}
to satisfy this latter goal, while \Par{imp} presents a simple numerical method 
by which the match to the target functions can be vastly improved.

\subsection{How Well do the DFs Re-Produce their Targets?} \label{sec:comp}
\ifpreprint\noindent\fi
We now specify the target functions $\Sigma(R)$ and $\sigma_R(R)$ and compare
these with the numerically evaluated velocity moments resulting from the 
various forms of warm-disk DFs. We restrict ourselves to exponential 
surface-density and velocity-dispersion profiles
\ben
	\Sigma(R)   &=& \Sigma_0\; {\rm e}^{-R/R_s}	\\
	\sigma_R(R) &=& \sigma_0\; {\rm e}^{-R/R_\sigma}
\een
with scale radii $R_s$ and $R_\sigma$. The gravitational potential is assumed
to be a power-law with circular speed curve $v_c\eq v_0(R/R_s)^\beta$ (cf.\ 
Appendix B of paper~I). Furthermore, we only consider models with $R_\sigma\eq3
R_s$ resulting in $\sigma_R^3\nc\propto\Sigma$ corresponding roughly to the 
situation in the Milky-Way disk. For flat ($\beta\eq0$), slightly rising 
($\beta\eq0.2$) or falling ($\beta\eq-0.2$) rotation curves, and for the four 
warm-disk DFs introduced in the last section, \Fig{comp} plots the logarithmic 
deviations of surface density and velocity dispersion for $\sigma_0\eq0.5v_0$ 
(\fig{comp}a-c), corresponding to $\sigma(R_0\ap2.5R_\Sigma)\ap0.2v_0$ 
reminiscent of the old-stellar disk of the Milky Way, and doubly as much, 
$\sigma_0\eq v_0$, for $\beta\eq0$ only (\fig{comp}d). 

\placefigure{fig:comp}

At $R\nc\ga R_\Sigma$ in all four cases, Shu's DF shows the strongest deviation 
both from the target surface density and velocity dispersion, while $\fnew$ 
gives the best match. On the other hand, at small radii $\fnew$, and to a 
smaller extent $\fa$, creates strongly rising $\Sigma_f$ and $\sigma_{R,f}$. 
The likely cause of this effect are eccentric orbits contributing significantly
to the surface density near their peri-centers, i.e.\ at small radii. 
In most applications, however, this should be no problem, since (i) there is 
negligibly little mass in these artificial cusps, and (ii) the central parts of
disk galaxies are dominated by dynamically hot bulges. Moreover, the technique
described in \Par{imp} below may be used to suppress these artificial cusps.

The case with an target of $\sigma_0\eq v_0$ (\fig{comp}d) clearly represents 
more a hot than a cold disk ($\sigma_R^2\pl\sigma_\phi^2\nc>\ov{v}^2_\phi$ out 
to $R\nc\simeq1.3R_\Sigma$ for $\fnew$), while the DFs were only meant to work 
for moderately warm disks (i.e.\ ordered motions dominating). Nonetheless, the 
moments created by $\fnew$ match the target functions reasonably well. The 
other two new warm-disk DFs, $\fa$ and $\fb$, are less useful than $\fnew$. 

From \Fig{comp}, we see that the DFs are much better in re-producing their 
target functions for rising than for falling rotation curves. Presumably, 
this is because for a fixed $\sigma_R(R)$, the ratio $\sigma_R/v_c$ is, at
most radii, smaller for rising than for falling rotation curves. A smaller
ratio $\sigma_R/v_c$ implies that the disk is colder and the DFs, which are
exact only in the cold limits, give a better match. 

\subsection{Improving the distribution functions} \label{sec:imp}
\ifpreprint\noindent\fi
The results of the last subsection show that DFs with $\sigma_R\nc\neq0$ only
approximately re-produce their target surface density and velocity dispersion.
The goal of this section is to get a much better match with some given surface 
density $\Sigma(R)$ and velocity dispersion $\sigma_R(R)$, by distinguishing
between the targets and the parameter functions that enter the functional form 
of the DF. That is, we replace the functions $\Sigma$ and $\sigma_R$ in the 
definitions of the warm-disk DFs with $\Sigma^\pr$ and $\sigma_R^\pr$, which 
are to be constructed such that the resulting moments of the DF closely match 
the target: $\Sigma_f\ap\Sigma\neq\Sigma^\pr$ and $\sigma_{R,f}\ap\sigma_R\neq
\sigma_R^\pr$.

\placefigure{fig:iter}
\placefigure{fig:iter-two}

\subsubsection{A Simple Algorithm to Construct $\Sigma^\pr$ and $\sigma_R^\pr$}
\label{sec:algol} \ifpreprint\noindent\fi
The simplest technique to optimize $\Sigma^\pr$ and $\sigma_R^\pr$ is to start 
with $\Sigma^\pr\eq\Sigma$ and $\sigma_R^\pr\eq\sigma$ and iterate the 
following steps.

\newcounter{lll}
\begin{list}{\arabic{lll}.}{\usecounter{lll} 
\leftmargin4mm \labelwidth3mm \labelsep1mm \itemsep0mm
\itemsep5pt plus 0.1pt
\parsep0pt plus 0.1pt
\topsep5pt plus 0.1pt}
\item	Compute $\Sigma_f$ and $\sigma_{R,f}$ due to the DF.
\item	Multiply $\Sigma^\pr$ with $\Sigma/\Sigma_f$ and $\sigma_R^\pr$ with 
	$\sigma_R/\sigma_{R,f}$.
\end{list}
\Fig{iter} shows the logarithmic deviations between target and actual $\Sigma$ 
and $\sigma_R$ that have been obtained after up to four iterations for the same 
stellar disk and potential as in \fig{comp}a. There is clearly a dramatic 
improvement already after one iteration: the average relative deviation has 
dropped by up to a factor ten. \Fig{iter-two} is similar to \fig{iter}, except 
that the target $\sigma_R$ is twice as large (the model used in \fig{comp}d) 
such that random motions are actually dominant inside about 1.3$R_\Sigma$. This 
time, the improvements are modest, and only in the case of $\fnew$ do the 
iterations reduce the deviation from the targets to a few per cent. 

\ifpreprint
  \begin{figure*}
	\centerline{\epsfxsize=160mm \epsfbox[32 315 582 709]{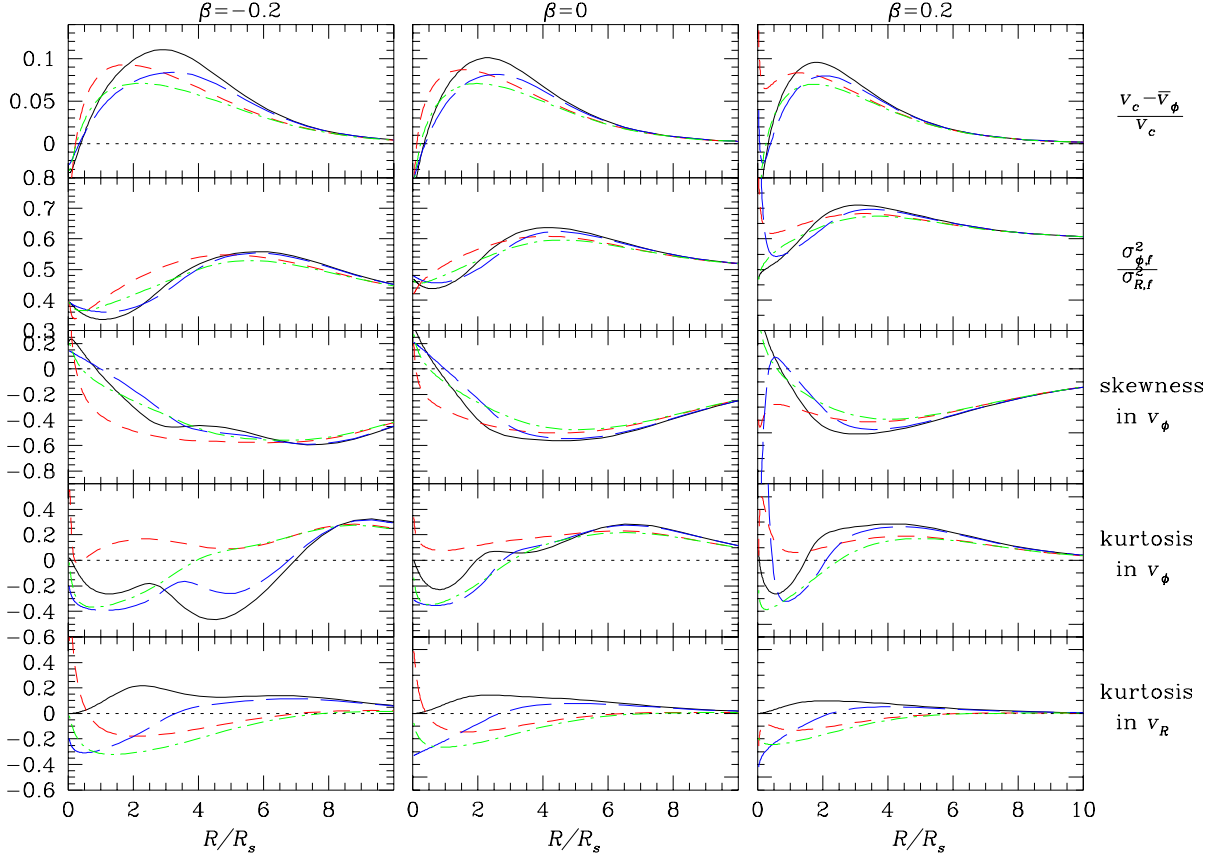}}
	\caption[]{\footnotesize
	Kinematics vs.\ $R$ for the warm-disk DFs $\fshu$ (solid), $\fnew$ 
	(dashed), $\fa$ (dot-dashed), and $\fb$ (long dashed) for an exponential
	disk with $R_\sigma\eq3R_\Sigma$ and $\sigma_0\eq0.5v_0$ in a slightly 
	falling ($\beta\eq{-}0.2$), flat ($\beta\eq0$), and slightly rising 
	($\beta\eq0.2$) rotation curve (same as in \fig{comp}a,b,c).
	\label{fig:kinem} }
  \end{figure*}

  \begin{figure*}
	\centerline{\epsfxsize=160mm \epsfbox[32 315 582 709]{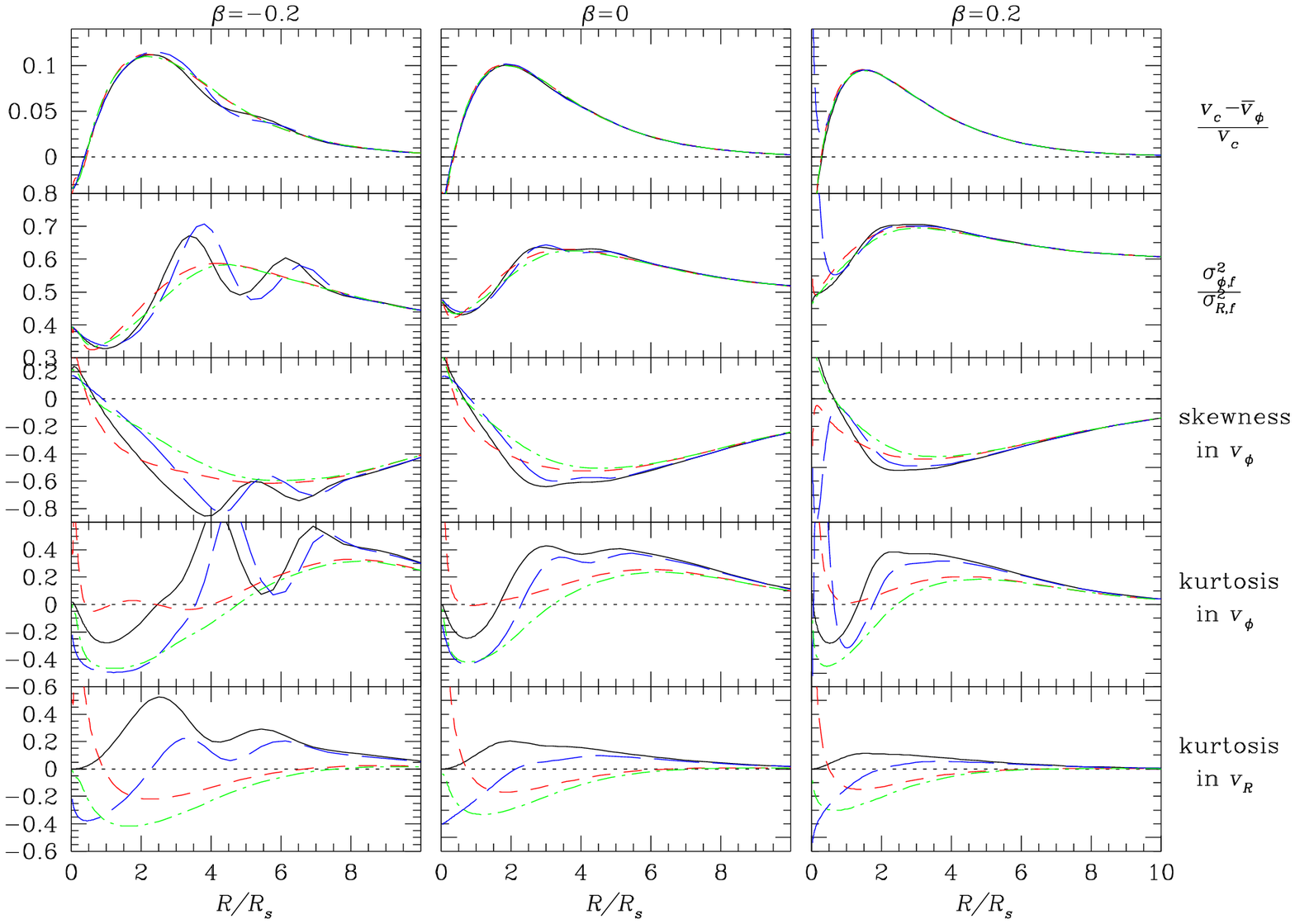}}
	\caption[]{\footnotesize
	Same as \Fig{kinem} but for DFs after four iterations of the algogithm
	of \Par{algol}. \label{fig:kinem-iter} }
  \end{figure*}
\fi

It seems, that after a few iterations no further improvement is achieved, i.e.\ 
the algorithm does converge to a model that definitely deviates from the target 
functions. For modest velocity dispersions this deviation is less than about one
per cent, but up to a few per cent or even more, depending on the DF, for 
stellar disks with significant random motions. From our explanation for the 
algorithm below, this is not astonishing, since the algorithm is very crude 
indeed; on the contrary, it is suprising that it works so well. 

\subsubsection{Why Does This Algorithm Work?} \ifpreprint\noindent\fi
The simple algorithm given above can be derived as follows. The integrations 
over velocity space that yield the moments may be cast into the form
\bsq \label{project} \ben
	\Sigma_f(R)     &=& \int \Sigma^\pr(R_1)\; K_\Sigma(R|R_1)\,\D R_1, \\
	\mu_f(R) 	&=& \int \mu^\pr(R_1)\; K_\mu(R|R_1)\,\D R_1
\een \esq
with $\mu\id\sigma_R^2\Sigma$. The kernels $K_\Sigma$ and $K_\mu$ are implicitly
defined by the form of the DF, but cannot be obtained explicitly. However, 
epicycle theory tells us that these Kernels should have a typical width of 
order $\sigma_R^\pr(R_1)/\kappa(R_1)$, which in turn is of the order 
$(\sigma_R^\pr/v_c)_{R_1}R_1$. Equation (\ref{project}) defines an inverse 
problem for finding the primed quantities given the unprimed ones. Clearly, to 
yield a physical meaningful DF, the primed quantites must not become negative, 
and a simple technique to solve for $\Sigma^\pr$ and $\sigma_R^\pr$ whilst 
ensuring positive definiteness is the Richardson-Lucy algorithm. This consists
of iterating on (\ref{project}) and the correction step
\bsq \label{correct} \ben
	\Sigma^\pr(R_1)	&\rightarrow& \Sigma^\pr(R_1) \times
		\int {\Sigma(R)\over\Sigma_f(R)} \,K_\Sigma(R|R_1)\,\D R,\\
	\mu^\pr(R_1) &\rightarrow& \mu^\pr(R_1) \times
		\int {\mu(R)\over\mu_f^2(R)} \,K_\mu(R|R_1)\,\D R.
\een \esq
The algorithm given above emerges when replacing the kernels in (\ref{correct}) 
with $\delta$-functions. Clearly, this is a very crude simplification, but 
computing the integrals in (\ref{correct}) is non-trivial, and furthermore, the 
scheme works reasonably well, as demonstrated above.

\ifpreprint	\section{T\fL{he} K\fL{inematics}}\label{sec:kinem}\noindent
\else		\section{The Kinematics} \label{sec:kinem}
\fi
As mentioned in the introduction, the distribution function $f(E,L)$ for a 
stellar disk is not uniquely specified by its surface density, velocity 
dispersion, and the gravitational potential. Rather, there is an infinite 
number of possible distribution functions with identical $\Sigma_f$ and 
$\sigma_{R,f}$, but with differences in, e.g., other velocity moments.
Here, we will study the kinematics, i.e. the velocity moments up to order 
four, that emerge from the various warm-disk DFs when applied to the same
stellar disk.

\placefigure{fig:kinem}
\placefigure{fig:kinem-iter}

For the case of an exponential disk with $R_\sigma\eq3R_\Sigma$ and $\sigma_0
\eq0.5v_0$ (as in \fig{comp}a-c), \Fig{kinem} plots various kinematical 
quantities resulting from the four different DFs of \Par{warm-dfs}.
\Fig{kinem-iter} differs only in that the DFs are those obtained after four 
iterations of the algorithm of \Par{algol}. Obviously, in some cases this 
rather crude algorithm tends to create DFs with somewhat``ragged'' kinematics.

The top panels show the run of the asymmetric drift velocity, which usually
peaks between 1 and 2 $R_\Sigma$ for these models. The iterated DFs show
almost identical behavior in their asymmetric drifts.

The second panels from top show the ratio between the kinetic energies in 
azimuthal and radial random motions, which in the solar neighborhood is 
measured to be about 0.42 (cf. Dehnen \& Binney 1998). In the limit of small 
velocity dispersions, we know from Oort's work that 
\beq \label{oort}
	\lim_{\sigma\to0}\; {\sigma_\phi^2\over\sigma_R^2} = {1\over2}
	\left[1+{\D\ln v_c\over\D\ln R}\right] = {-B\over A-B} = \gamma^{-2},
\eeq
where $A$ and $B$ are Oort's constants. Indeed, at large $R$, this value is 
reached by the moments of the DFs. However, at radii $R\nc\ga2R_\Sigma$, all 
four DFs predict $\sigma_\phi^2/\sigma_R^2$ to be significantly larger than 
this limit. This is a known result for exponential disks\footnote{
	This fact is not very well known, though, and it cannot be stressed
	enough that Oort's equation (\ref{oort}) is correct only in that limit 
	and must not be applied to the situation in the solar neighborhood, 
	even though this has been done many times in the past.}
(cf.\ \cite{ec93,cb94}). As the third panels from top of Figures \ref{fig:kinem}
and \ref{fig:kinem-iter} show, the $v_\phi$ distributions are always signicantly
skewed in the sense that they have a low-$v_\phi$ tail. This tail is caused by 
stars originating from $R_1\nc<R$ with large eccentricities -- because of the 
exponential $\Sigma(R)$ and $\sigma_R(R)$ relations, there are many fewer stars 
with large eccentricity originating from $R_1\nc>R$, which would create a 
high-$v_\phi$ tail. Such a skewness in the azimuthal motions is indeed observed 
in the solar neighborhood, cf.\ Dehnen (1998).

It is only in the kurtoses, i.e.\ moments of the fourth order, that the four
DFs appear to differ significantly. While the kurtosis for the azimuthal
motion is mostly positive, that of the radial motion may have either sign.

\ifpreprint	\section{S\fL{ampling of} P\fL{hase}-S\fL{pace} P\fL{oints}}
		\label{sec:sample} \noindent
\else		\section{Sampling of Phase-Space Points}
		\label{sec:sample}
\fi
For $N$-body simulations and similar studies, it is of importance to draw
phase-space points $(R,\dot{R},\phi,\dot{\phi})_i,\,i=0,\dots N$ from
the distribution function. One is often interested not in randomly distributed
phase-space points but in a more regular distribution, where the points are
equidistant in the phases (angle variables). Such a distribution has minimal
noise, a fact important in numerical studies of stability properties (``quiet 
start'' technique). 

From the general form (\ref{f-motiv}) of the warm-disk DFs presented, it is
straightforward to derive a procedure for sampling $(E,L)$ as follows. 
First, determine $R_1$ according to $\Sigma(R)$, and translate it to either
$E\eq E_c(R_1)$ or $L\eq L_c(R_1)$ according to the functional form of the DF.
Second, sample the exponential in radial action and compute the remaining
integral: $E$ or $L$. Here is a general algorithm that allows for parameter 
functions $\Sigma^\pr$ and $\sigma^\pr$ different from the targets and 
accurately corresponds to the DFs of \Par{warm-dfs}.

\setcounter{lll}{0}
\begin{list}{\arabic{lll}.}{\usecounter{lll} \leftmargin4mm 
\itemsep0pt plus 1pt
\parsep0pt plus 1pt
\topsep2pt plus 1pt
\labelwidth3mm \labelsep1mm \itemsep2ex}
\item 	Choose a radius $R$ randomly from the cumulative distribution
	\beq 
		P(R) = \int_0^R R^\pr\,\Sigma(R^\pr)\,\D R^\pr
	\eeq
	and determine $L\eq L_c(R)$ for $\fshu$ and $\fb$ or $E\eq E_c(R)$ for 
	$\fnew$ and $\fa$. Evaluate the correction factor $g_1=\Sigma^\pr(R)/
	\Sigma(R)$.

	In the case of $\fb$, the sign of $L$ must be chosen to be positive or
	negative with relative probabilities 
	\[ 
		1:\exp\big[-2\Omega(R_1)|L|/\sigma^2_R(R_1) \big].
	\]
\item	Choose $\xi\in(0,1)$ randomly and determine
	\beq \label{sam-Itwo} \bea{r@{\,=\,}l@{\qquad{\rm for}\;\;}l}
	   E & E_c(L) - \sigma^{\pr2}_R(R)\ln\xi 		& \fshu,\\[1ex]
	   L & L_c(E) + \sigma^{\pr2}_R(R)\ln\xi\,/\Omega(R)	& \fnew,\\[1ex]
	   L & L_c\!\big(E+\sigma^{\pr2}_R(R)\ln\xi\big) 	& \fa,	\\[1ex]
	   E & E_c\!\big(|L|-\sigma^{\pr2}_R(R)\ln\xi\,/\Omega(R)\big) & \fb.
	\eea \eeq
	If $E\not\in[\Phi(0),\Phi(\infty)]$ or $L\not\in[-L_c(E),L_c(E)]$,
	go back to step 1 for another try.
\item	Integrate the orbit with these values of $(E,L)$ over one radial 
	period $T_R$, compute $\omega_R(E,L)\equiv2\pi/T_R$, and evaluate the 
	correction factor $g_2\eq\kappa(R)/\omega_R(E,L)$.
\item 	Sample $N_{\rm sam}$ phase-space points randomly 
	from $\phi\in[0,2\pi)$ and $t\in[0, T_R)$, use a table made during orbit
	integration to obtain $R(t)$ and $\dot{R}(t)$; $\dot{\phi}\eq L/R^2$. 
	Here $N_{\rm sam}$ is either of the two integers next to $g_1\,g_2\,
	N_{\rm orb}$, chosen with probabilities such that the mean is equal to
	$g_1\,g_2\,N_{\rm orb}$.
	
\item	Iterate steps 1 to 4 until the desired number $N$ of points is sampled.
\end{list}
Here, $N_{\rm orb}$ is, approximately, the number of points chosen per orbit, 
and is usually between 1 and $N^{1/2}$. If noise is to be minimized, e.g.\ if 
initial conditions for a quiet start are needed, use $N_{\rm orb}\nc\gg1$ and 
quasi-random numbers or even equidistant points, otherwise pseudo-random 
numbers. If only $(E,L)$ shall be sampled, set $N_{\rm orb}\eq1$. Note that for
$\fshu$ and $\fb$ the correction factor $g_2$ can become quite large for 
eccentric orbits, since these DFs are based on estimating $\omega_R$ by 
$\kappa(L)$, which goes to infinity as $L\to0$ in centrally non-harmonic 
potentials, as, e.g., in the Milky Way. In the case of $\fb$, one first 
determines $|L|$ and has to choose the sign afterwards. 

I have checked the correctness of the above algorithm numerically by computing 
$\Sigma_f$ and $\sigma_{Rf}$ from $10^6$ sampled phase-space points: they
agreed within their statistical uncertainties with the results from direct
numerical integration.

\ifpreprint	\section{C\fL{onclusion}} \label{sec:sum}\noindent
\else		\section{Conclusion} \label{sec:sum}
\fi
\subsection{Summary} \ifpreprint\noindent\fi
Distribution functions for dynamically warm stellar disks may be considered
as warmed-up versions of completely cold disks, in which all stars rotate on
circular orbits. In fact, this picture for the DF of a stellar disk may be
considered a good description for its formation history. In \Sec{warm-dfs},
I presented four different functional forms for warm-disk DFs $f(E,L)$, given 
the gravitational potential $\Phi(R)$ and prescriptions for the surface-density
and velocity-dispersion profiles, $\Sigma(R)$ and $\sigma_R(R)$. One of them,
$\fshu$, was introduced by Shu in 1969 and has already been used in studies, 
e.g.\ of the solar neighborhood kinematics (see Bienaym\'e 1999 for a recent 
example). The other three DFs have not been described in the literature so far. 

In \Sec{assess}, the surface-density and velocity-dis\-per\-sion profiles 
produced by the DFs as moments were compared to the target profiles. The 
comparison shows that, for a typical case reminiscent to the old stellar disk 
in the Milky Way, Shu's DF, $\fshu$ results in the largest deviations between
moments and targets, while the new DF (\ref{fnew}) gives deviations that are 
about three times smaller. This new DF even gives an useful model when the 
stellar disk is dynamically hot within about one scale radius. Actually, the 
deviation of the moments from the target profiles $\Sigma(R)$ and $\sigma_R(R)$ 
may be considerably reduced by allowing the corresponding parameter functions 
appearing in the definition of the DFs to differ from these targets. In 
\Sec{imp}, a simple algorithm for this purpose is presented, which after only 
four iteration reduces the deviations in $\Sigma$ and $\sigma_R$ to well below 
1 per cent for a typical case. Again, the DF $\fnew$ is best suited for this 
game, as it leads to the smallest deviations also in the iterated case. Another 
advantage of $\fnew$ is that is naturally extends to negative $L$, while 
$\fshu$ make the rather unnatural assumption that there are no stars with 
$L\nc<0$.

In \Sec{kinem}, I have compared the velocity moments up to fourth order of the 
four DFs iterated and non-iterated. For all four DFs, the ratio of azimuthal to 
radial velocity dispersion squared, $\sigma^2_\phi/\sigma^2_R$ is significantly 
{\em larger\/} than $\half[1\pl \D\ln v_c/\D\ln R]$ for $R$ larger than about 2 
disk scale radii. The value $\half[1\pl \D\ln v_c/\D\ln R]$ is expected from 
Oort's \eqn{oort}, which in turn is based on the epicycle approximation whereby 
ignoring the strong gradients in an exponential disk. In the solar neighborhood,
$\sigma^2_\phi/\sigma^2_R\approx0.42$ is observed, i.e.\ {\em smaller\/} than 
what we expect for a near-flat rotation curve. This deviation must be caused by 
a wrong assumption in the models, most likely that of axisymmetry -- an 
investigation in terms of the influence of the Galactic bar is the subject of 
a future paper.

Finally, in \Sec{sample}, I give an algorithm for the sampling of phase-space
points $(R,p_R,\phi,L)$ from any one of the four DFs. This should be 
particularly useful for the creation of initial conditions for $N$-body 
simulations, either of the evolution of disk galaxies (bar formation, warps, 
etc.) or of mergers involving disk galaxies.

\subsection{Thick Disks} \ifpreprint\noindent\fi
An obvious extension of the thin-disk DFs presented in this paper are DFs 
describing stellar disks with some finite thickness. For this purpose, one needs
a handle on the vertical motion of the stellar orbits. Unfortunately, typical  
three-dimensional axisymmetric potentials generally do not support a global 
third integral of motion $I_3$, even though most orbits are regular. There are 
various ways of different sophistication which can be used to proceed. First, 
the most basic and in principle only wholly correct procedure is Schwarzschild's
(1979) method of modeling a stellar system as superposition of individual 
orbits, which are obtained by numerical integration. Second, one may ignore the 
details of the phase-space structure and obtain a global third integral by 
perturbation techniques or torus fitting. In this case, (\ref{f-motiv}) can be 
generalized to
\beq
	f 	\approx{f_1(R_1)\over\sigma_R^2(R_1)\,\sigma_z^2(R_1)}\,
		\exp\!\left[-{\oR\JR\over\sigma^2_R(R_1)}
			    -{\omega_z J_z\over\sigma^2_z(R_1)}\right]
\eeq
with the vertical action $J_z$ (\cite{bin87,db96}). Third, one may use the 
vertical energy
\[
	E_z \equiv \case{1}{2} v_z^2 + \Phi(R,z) - \Phi(R,0)
\]
for $I_3$. This approach is presumably still very good for many applications, 
since $E_z$ is conserved to reasonable accuracy for many disk stars (but not 
for halo stars). Finally, one may restrict the analysis to a potential of 
St\"ackel form, such that a globally valid $I_3$ explicitly exists for all 
orbits (see Bienaym\'e, 1999, for a recent application of this technique).

\subsection{Computer Programs Available} \ifpreprint\noindent\fi
The computer programs that I have designed in the course of this study
are written in {\sf C++} and contain the following features.

\setcounter{lll}{0}
\begin{list}{\arabic{lll}.}{\usecounter{lll} \leftmargin4mm 
\itemsep0pt plus 0.1pt
\parsep0pt plus 0.1pt
\topsep3pt plus 0.1pt
\labelwidth3mm \labelsep1mm}
\item 	A general concept for one-dimensional potentials and several 
	implementations (power-law, $\gamma$-models, Iso\-chro\-ne) are given. 
	Among other things, this contains routines for numerical orbit 
	integration and evaluation of $J_R$, $\omega_R$, and $\omega_\phi$.
\item	A general concept of warm-disk DFs of the form (\ref{f-motiv})
	is given, which allows for integration of general velocity moments 
	$\ov{\Sigma v_R^m v_\phi^n}\id\!\int\D^2\!\B{v}\,v_R^m v_\phi^n f$. 
	It also enables computation of $f(E,L)$, $R_1(E,L)$, and the 
	derivatives of $f$ w.r.t.\ the parameter functions $\Sigma^\pr$ and 
	$\sigma_R^\pr$ (useful for fitting $f$ to some data).
\item	Code for sampling phase-space points from a DF either pseudo- or 
	quasi-randomly (the latter to reduce noise) according to the algorithm 
	given in \Sec{sample}.
\item	The iterative algorithm of \Sec{algol} to improve the match between
	moments and targets is implemented.
\end{list}
These programs are electronically available from me upon request. There is
also an interface which allows one to use a reduced version of these routines
from {\sf C} programs.

%
%

\ifpreprint
  \def\thebibliography#1{\subsection*{R\fL{eferences}}
    \list{\null}{\leftmargin 1.2em\labelwidth0pt\labelsep0pt\itemindent -1.2em
    \itemsep0pt plus 0.1pt
    \parsep0pt plus 0.1pt
    \parskip0pt plus 0.1pt
    \usecounter{enumi}}
    \def\refpar{\relax}
    \def\newblock{\hskip .11em plus .33em minus .07em}
    \sloppy\clubpenalty4000\widowpenalty4000
    \sfcode`\.=1000\relax}
  \def\endthebibliography{\endlist}
\fi

%
%

\ifpreprint \relax \else
\clearpage \onecolumn

\begin{figure}\caption[]{
	Logarithmic deviations between the target surface density and velocity 
	dispersion ($\Sigma$, $\sigma_R$) and those actually created by the 
	warm-disk DF ($\Sigma_f$, $\sigma_{R,f}$) in power-law potentials
	with circular speed $v_c\nc\propto R^\beta$. Solid lines are for 
	$\fshu$, dashed for $\fnew$, dot-dash for $\fa$, and long-dashed for 
	$\fb$. In all four sub-figures, both $\Sigma$ and $\sigma_R$ decay
	exponentially with scale lengths $R_\sigma\eq3R_s$. In (a), (b), 
	and (c), the central velocity dispersion is $0.5v_0$, while it is twice 
	as large for (d). The rotation curve is flat in (a) and (d), slightly
	falling in (b) and slightly rising in (c). Notice the different scales
	of the $y$-axes. \label{fig:comp} }
\end{figure}

\begin{figure}\caption[]{
	The same as in \fig{comp}a, but for the DFs $\fshu$ (a) and $\fnew$ (b)
	after none (thin), one (dotted), two (dashed), three (long-dashed), and 
	four (bold dot-dash) iterations of the scheme of \Sec{algol}. Notice 
	the different scales on the $y$-axes. \label{fig:iter} }
\end{figure}

\begin{figure}\caption[]{
	As \fig{iter} but for the stellar disk and rotation curve as in 
	\fig{comp}d (which is hot inside $\sim1.3R_\Sigma$) and after none 
	(thin), one (dotted), two (dashed), five (long-dashed), and seven (bold 
	dot-dash) iterations of the scheme of \Sec{algol}. Notice the different 
	scales on the $y$-axes. \label{fig:iter-two} }
\end{figure}

\begin{figure}\caption[]{
	Kinematics vs.\ $R$ for the warm-disk DFs $\fshu$ (solid), $\fnew$ 
	(dashed), $\fa$ (dot-dashed), and $\fb$ (long dashed) for an exponential
	disk with $R_\sigma\eq3R_\Sigma$ and $\sigma_0\eq0.5v_0$ in a slightly 
	falling ($\beta\eq{-}0.2$), flat ($\beta\eq0$), and slightly rising 
	($\beta\eq0.2$) rotation curve (same as in \fig{comp}a,b,c).
	\label{fig:kinem} }
\end{figure}

\begin{figure}\caption[]{
	Same as \Fig{kinem} but for DFs after four iterations of the algogithm
	of \Sec{algol}. \label{fig:kinem-iter} }
\end{figure}

%
%

	\clearpage
	\centerline{ \epsfxsize=150mm\epsfbox[27 325 590 716]{Dehnen2.fig1.ps}}
        \centerline{Fig.~\ref{fig:comp}}

	\clearpage
	\centerline{ \epsfxsize=75mm\epsfbox[25 325 303 716]{Dehnen2.fig2.ps} }
        \centerline{Fig.~\ref{fig:iter}}

	\vspace{2cm}
	\centerline{ \epsfxsize=75mm\epsfbox[25 325 303 716]{Dehnen2.fig3.ps} }
        \centerline{Fig.~\ref{fig:iter-two}}

	\clearpage
	\centerline{\epsfxsize=160mm \epsfbox[32 315 582 709]{Dehnen2.fig4.ps}}
        \centerline{Fig.~\ref{fig:kinem}}

	\clearpage
	\centerline{\epsfxsize=160mm \epsfbox[32 315 582 709]{Dehnen2.fig5.ps}}
        \centerline{Fig.~\ref{fig:kinem-iter}}

\fi 
\end{document}